\documentstyle[e-e-ijmpa,twoside]{article}
\input psfig
\input epsfig.sty
\def\nn{\noindent}
\def\Re{{\cal R \mskip-4mu \lower.1ex \hbox{\it e}\,}}
\def\Im{{\cal I \mskip-5mu \lower.1ex \hbox{\it m}\,}}
\def\ie{{\it i.e.}}

\def\etal{{\it et al.}}

\def\to{\rightarrow}

\def\subw{_{\rm w}}
\def\mh{\ifmmode m\sbl H \else $m\sbl H$\fi}
\def\mch{\ifmmode m_{H^\pm} \else $m_{H^\pm}$\fi}
\def\mt{\ifmmode m_t\else $m_t$\fi}
\def\mc{\ifmmode m_c\else $m_c$\fi}
\def\mz{\ifmmode M_Z\else $M_Z$\fi}
\def\mw{\ifmmode M_W\else $M_W$\fi}
\def\mws{\ifmmode M_W^2 \else $M_W^2$\fi}
\def\mhs{\ifmmode m_H^2 \else $m_H^2$\fi}   
\def\mzs{\ifmmode M_Z^2 \else $M_Z^2$\fi}
\def\mts{\ifmmode m_t^2 \else $m_t^2$\fi}
\def\mcs{\ifmmode m_c^2 \else $m_c^2$\fi}
\def\mchs{\ifmmode m_{H^\pm}^2 \else $m_{H^\pm}^2$\fi}
\def\ztwo{\ifmmode Z_2\else $Z_2$\fi}
\def\zone{\ifmmode Z_1\else $Z_1$\fi}
\def\mtwo{\ifmmode M_2\else $M_2$\fi}
\def\mone{\ifmmode M_1\else $M_1$\fi}
\def\tb{\ifmmode \tan\beta \else $\tan\beta$\fi}
\def\xw{\ifmmode x\subw\else $x\subw$\fi}
\def\ch{\ifmmode H^\pm \else $H^\pm$\fi}
\def\lum{\ifmmode {\cal L}\else ${\cal L}$\fi}
\def\inpb{\ifmmode {\rm pb}^{-1}\else ${\rm pb}^{-1}$\fi}
\def\infb{\ifmmode {\rm fb}^{-1}\else ${\rm fb}^{-1}$\fi}
\def\epem{\ifmmode e^+e^-\else $e^+e^-$\fi}
\def\gg{\ifmmode \gamma\gamma\else $\gamma\gamma$\fi}
\def\ppb{\ifmmode \bar pp\else $\bar pp$\fi}
\def\bsg{\ifmmode B\to X_s\gamma\else $B\to X_s\gamma$\fi}
\def\bsll{\ifmmode B\to X_s\ell^+\ell^-\else $B\to X_s\ell^+\ell^-$\fi}
\def\bstt{\ifmmode B\to X_s\tau^+\tau^-\else $B\to X_s\tau^+\tau^-$\fi}

\newskip\zatskip \zatskip=0pt plus0pt minus0pt
\def\matth{\mathsurround=0pt}
\def\lsim{\mathrel{\mathpalette\atversim<}}

\def\atversim#1#2{\lower0.7ex\vbox{\baselineskip\zatskip\lineskip\zatskip
  \lineskiplimit 0pt\ialign{$\matth#1\hfil##\hfil$\crcr#2\crcr\sim\crcr}}}

\renewcommand{\thefootnote}{\fnsymbol{footnote}}	

\begin{document}


\rightline{\vbox{\halign{&#\hfil\cr
&SLAC-PUB-7775\cr
&March 1998\cr}}}

\normalsize\textlineskip
\thispagestyle{empty}

\title{{PROBING TOP-QUARK COUPLINGS AT LEPTON AND PHOTON COLLIDERS}
\footnote{To appear in the {\it Proceedings of the $2^{nd}$ International
Workshop on $e^-e^-$ Interactions at TeV Energies}, Santa Cruz, CA,
22-24 September 1997}
}

\author{ {JOANNE L. HEWETT}
\footnote{Work supported by the Department of Energy, 
Contract DE-AC03-76SF00515}
}

\address{Stanford Linear Accelerator Center,\\
Stanford University, Stanford, CA 94309, USA}

\maketitle\abstracts{The ability of high energy lepton and photon colliders 
to probe the gauge couplings of the top-quark is summarized.}

\setcounter{footnote}{0}
\renewcommand{\thefootnote}{\alph{footnote}}

\vspace*{1pt}\textlineskip	

\section{Introduction} 

The Standard Model (SM) has provided a remarkably successful description of
almost all available data involving the strong and electroweak interactions.
In particular, the discovery of the top-quark at the Tevatron with a 
mass\cite{topmass} $m_t=173.5\pm 5.2$ GeV, close to that anticipated by
fits to precision electroweak data\cite{ewwg} 
is a great triumph.  Nonetheless,
the SM leaves many fundamental questions unanswered, and this has led to 
numerous speculations on possible manifestations of new physics.  Since the
top-quark is the most massive fermion in the SM, it may provide a unique
insight to new interactions originating at a higher scale.  In fact, the
detailed physics of the top-quark may be significantly different from what
is predicted by the SM, making precision measurements of all its properties
mandatory.

One possible manifestation of new interactions in the top-quark sector is to 
alter its couplings to the SM gauge bosons, $W\,, Z\,,\gamma\,,$ or $g$.
This possibility, extended to all of the fermions of the third generation,
has attracted much attention\cite{htt} in the recent literature.  The
most general gauge invariant, non-renormalizable three-point $t\bar t\gamma\,,
Z\,, g$ interactions can be written as
\begin{equation}
{\cal L} = g_V\bar t\left[ \gamma_\mu(f_1^V+f_3^V\gamma_5) +
{i\sigma_{\mu\nu}q^\nu\over 2m_t}(f_2^V+f_4^V\gamma_5)\right] tV^\mu \,,
\label{lagrange}
\end{equation}
where $g_\gamma=e\,, g_Z=g/2c_w$, and $g_g=g_sT_a\,$ for $V^\mu=A^\mu\,,
Z^\mu$, and $G^{a\mu}$, respectively, and $q^\nu$ corresponds to the 
momentum carried by the gauge boson.  In the SM, 
\begin{eqnarray}
f_1^\gamma = Q_t\,, & & f_3^\gamma = 0\,, \nonumber\\
f_1^Z = v_t\,, & & f_3^Z = a_t\,, \\
f_1^g = 1\,, & & f_3^g = 0\,, \nonumber
\end{eqnarray}
with $Q_t\,, v_t\,, a_t$ being the top-quark's electric charge, and vector, 
axial-vector couplings to the $Z$.  At tree-level $f_{2,4}^V=0$.  The
form factor $f_2^V$ receives corrections of order $\alpha_s/\pi$ at loop-level,
while $f_4^V$ remains zero through 2-loops.  Gauge invariance will also lead
to new four-point interactions involving two gauge bosons and the top-quark.  In
most cases gauge invariance will relate any of the trilinear $t\bar t\gamma/Z$
anomalous couplings to others involving the $tbW$ vertex.  However,
Escribano and Masso\cite{escrib} have shown that in general all of the
anomalous three-point couplings involving the neutral gauge bosons can be
unrelated, even when the underlying operators are SM gauge invariant.  Of
course, within any particular new physics scenario the anomalous couplings
will no longer be independent.  We note that $f_{2,4}$ take the form of
dipole moment type couplings, with the magnetic dipole term $f_2$ being
CP conserving and $f_4$, the electric dipole moment term, violating CP.
These two form factors are also commonly denoted as $\kappa_V$ and 
$\tilde\kappa_V$ in the literature.

A generalization of new physics effects on these form factors can be written as
\begin{equation}
f_{1,3} \sim f^{SM}_{1,3}\left( 1+{q^2\over\Lambda^2}\right) \,, \quad\quad
f_{2,4}\sim {m_t^2\over\Lambda^2} \,,
\end{equation}
where $\Lambda$ represents the scale of the new interactions.  The 
importance of the large top mass is clear in this parameterization.
The $\gamma$ and $Z$ 
dipole moment form factors have been explicitly calculated in 
supersymmetry,\cite{bartl} where they are found to take on values typically
of order a few $\times 10^{-4}$ for $q^2$ values near $4M^2$ with $M$ being
the supersymmetric mass scale.  The one-loop corrections to the chromomagnetic
dipole form factors have recently been evaluated\cite{columbia} to be
$-0.004\leq f_2^g\leq -0.001$ in the SM and $-0.001\leq f_2^g\leq -0.01$ 
in Supersymmetry.

\section{Indirect Constraints}

\begin{figure}[t]
\centerline{
\psfig{figure=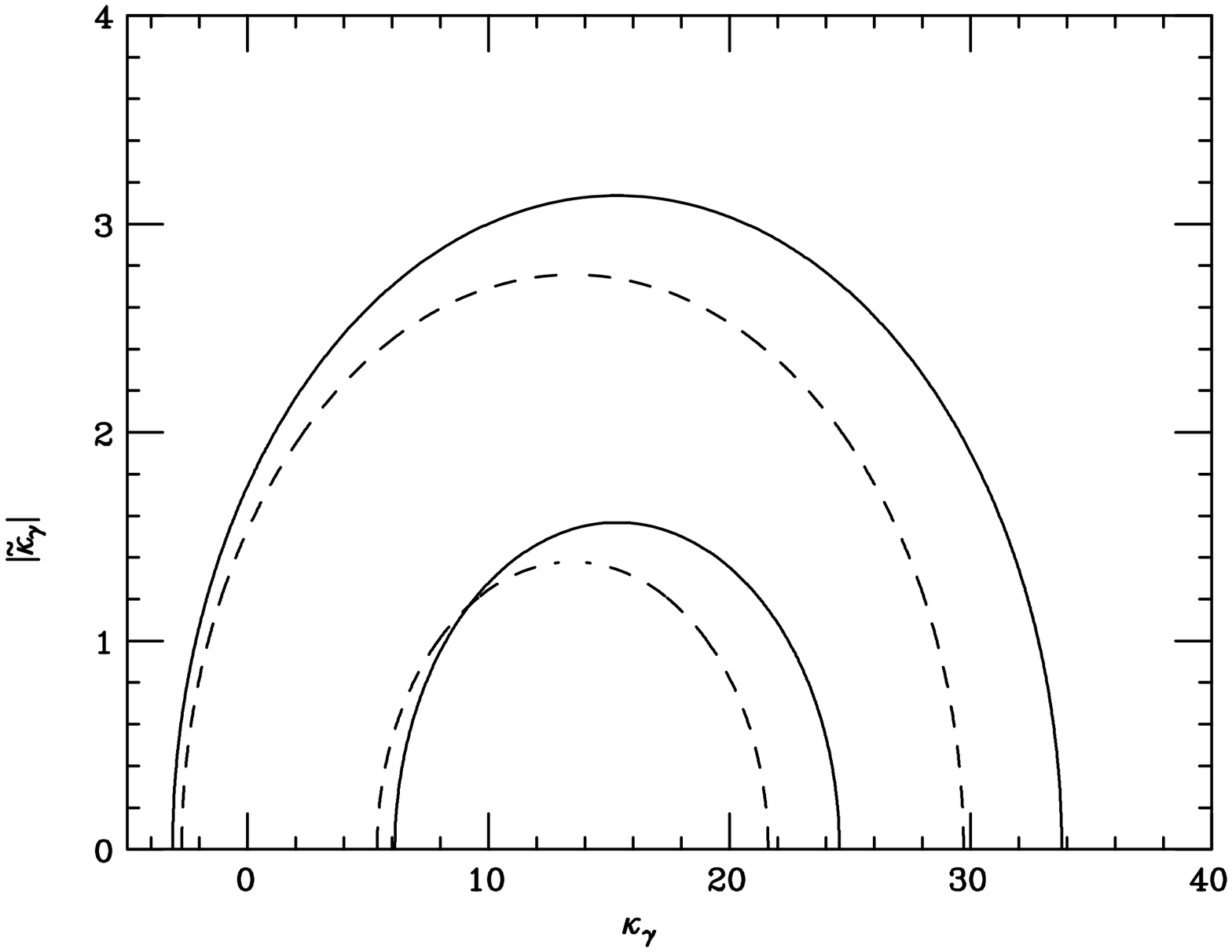,height=8cm,width=10cm,angle=-90}}
\vspace*{-0.5cm}
\fcaption{Constraints on anomalous top-quark photon couplings from $B\to X_s
\gamma$, assuming $m_t(m_t)=170$ GeV.
The solid and dashed curves correspond to the cases described in 
the text.  The allowed regions lie inside the semi-circles.} 
\label{bsg}
\end{figure}

Anomalous couplings of the top-quark to on-shell photons and gluons would
modify the rate for $B\to X_s\gamma$.  The presence of the magnetic and/or
electric dipole moment terms in Eq. (\ref{lagrange}) would affect the Wilson 
coefficients of the dipole operators which mediate $b\to s$ transitions by
\begin{equation}
C_{7,8}(M_W) = C_{7,8}^{SM}(M_W)+\kappa_{\gamma,g}F_{1_{7,8}}(m_t^2/M_W^2)
+\tilde\kappa_{\gamma,g}F_{2_{7,8}}(m_t^2/M_W^2)\,.
\end{equation}
$C_{7,8}^{SM}$ are the SM forms of the one-loop
matching conditions for the magnetic and chromomagnetic dipole operators 
${\cal O}_{7,8}$.  We remind the reader that the decay $B\to X_s\gamma$ is 
mediated by the magnetic dipole operator ${\cal O}_7$, however ${\cal O}_8$
also contributes indirectly via operator mixing.  The functions $F_{1,2}$ can be
found in Ref. \citenum{jlhtgr}.  Comparing the resulting branching fraction to
the CLEO measurement\cite{cleo} of the inclusive rate yields the bounds
in Fig. \ref{bsg}.  In this figure, the allowed region is given by the area
inside the solid (dashed) semi-circle when $\kappa_g,\tilde\kappa_g=0$
$(=\kappa_\gamma,\tilde\kappa_\gamma)$.

\vspace*{-0.5cm}
\nn
\vspace*{0.4mm}

Similar indirect constraints\cite{sally} can be found for $t\bar tZ$ 
couplings from the reaction $Z\to b\bar b$.

\section{Direct Probes at Colliders}

The direct production of top-quarks at high energy colliders provides the
best opportunity to probe these effective couplings.  It has been 
shown\cite{tgr} that non-standard $t\bar tg$ couplings can lead to significant 
modifications in the characteristics of top pair production at hadron 
colliders.  Since $t\bar t$ production at the Tevatron is dominated by
the invariant mass region near threshold, there is little effect on the
shape of the kinematic distributions from these couplings, whereas the total 
cross section can deviate substantially from SM expectations.  Current
cross section measurements\cite{topmass} hence constrain the chromomagnetic
dipole moment to $|f_2^g|\lsim 0.15-0.20$.  The larger partonic center of mass
energies at the LHC allow access beyond the threshold region
and hence yield much higher sensitivity to these couplings
via the $p_t$ and pair invariant mass ($M_{t\bar t}$) distributions.
The analysis of Rizzo\cite{tgrlhc} demonstrates that chromomagnetic dipole
moments of order $|f_2^g|\simeq 0.06$ can be probed at the LHC.

\begin{figure}[t]
\centerline{
\psfig{figure=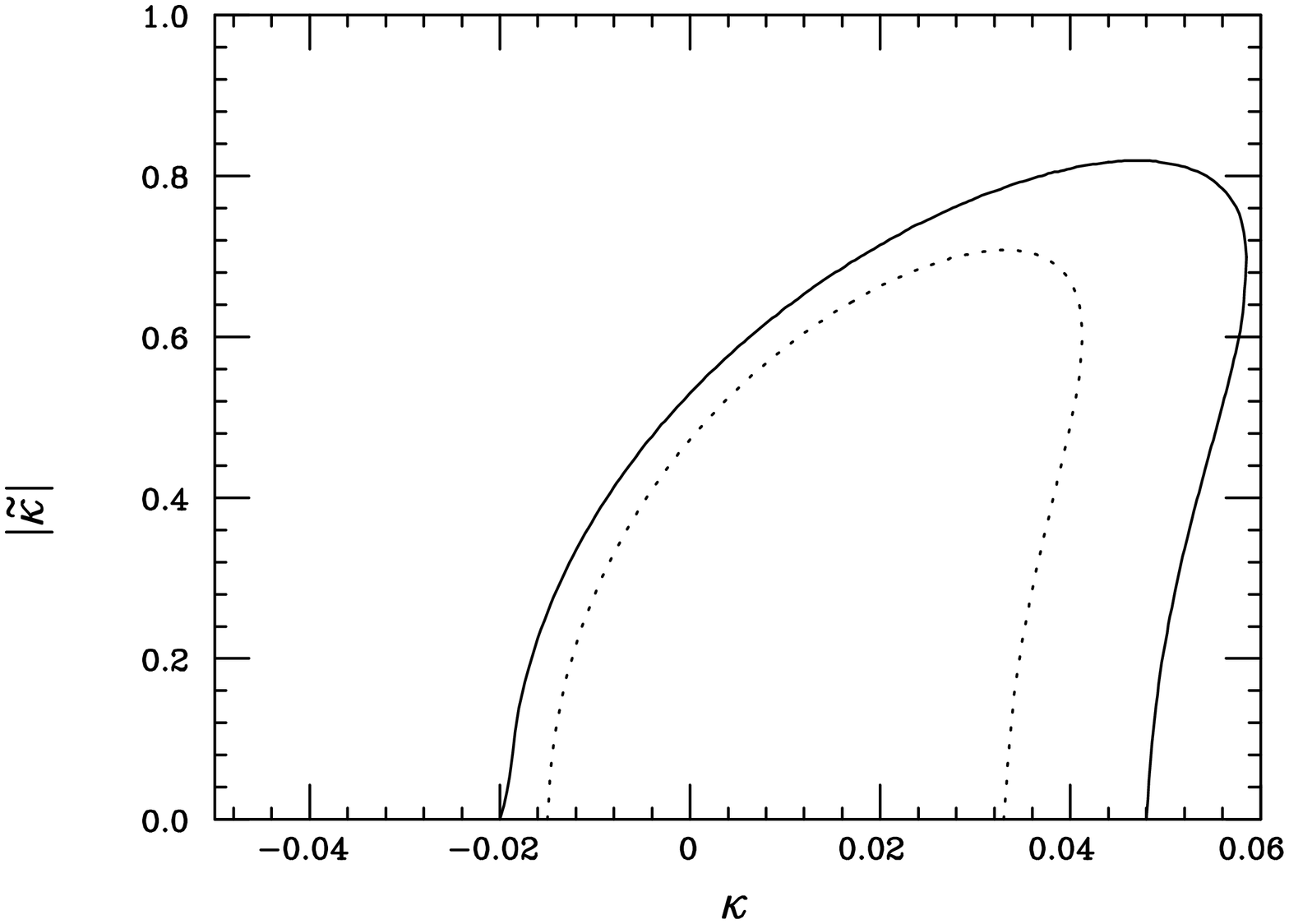,height=6.5cm,width=7cm,angle=-90}
\hspace*{-5mm}
\psfig{figure=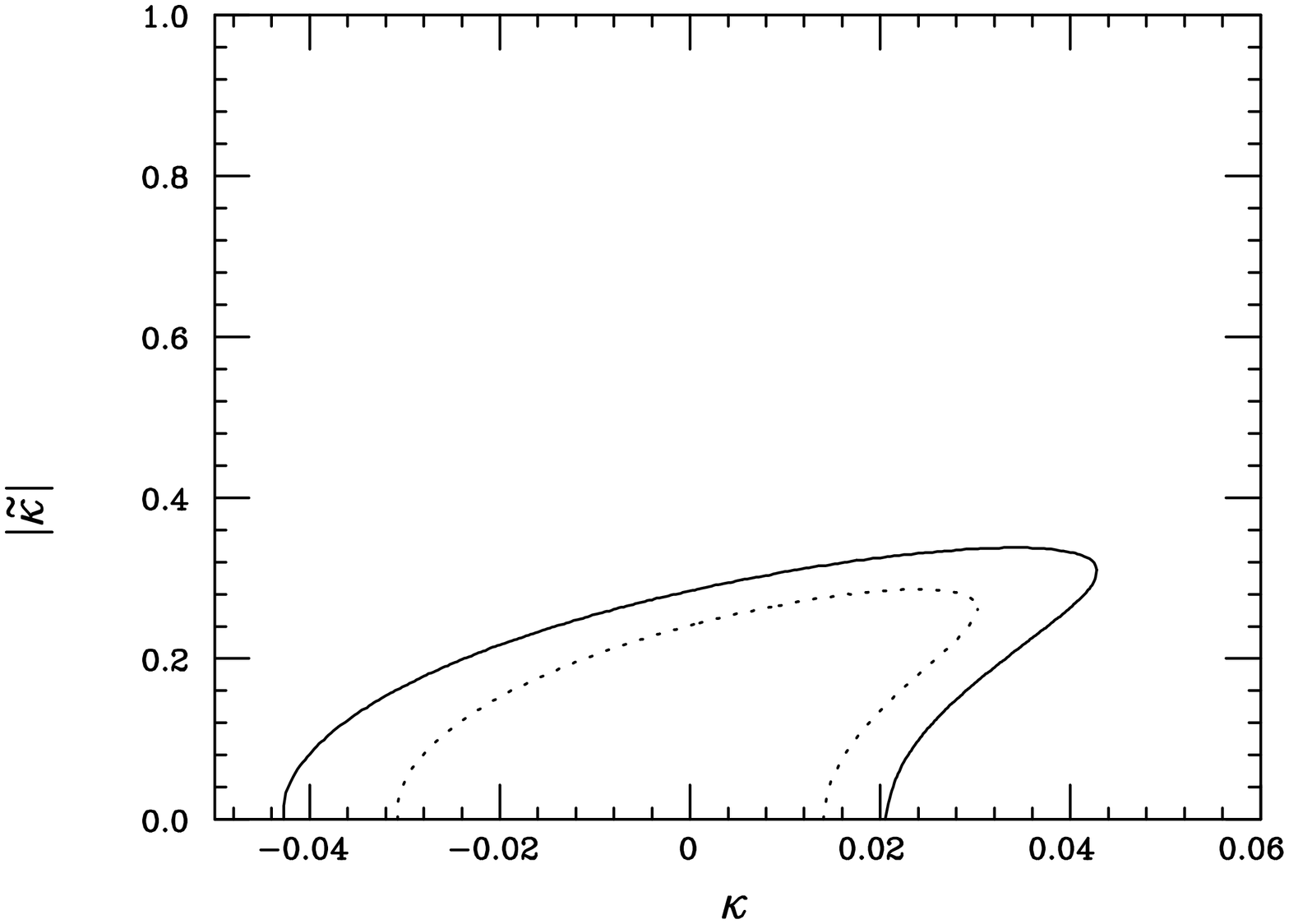,height=6.5cm,width=7cm,angle=-90}}
\vspace*{-0.5cm}
\fcaption{$95\%$ C.L. allowed region in the $f_4^g - f_2^g$ plane (denoted
here as $\tilde\kappa - \kappa$) from Ref. \citenum{tgrnlc}.  The left figure
employs $E_g^{min}=25 $ GeV at a 500 GeV NLC with a 50 (solid) or 100 (dotted)
fb$^{-1}$ data sample, while the right figure uses
employs $E_g^{min}=50 $ GeV at $\sqrt s=1$ TeV with 100 (solid) or 200 (dotted)
fb$^{-1}$.} 
\label{ttg}
\end{figure}

Anomalous chromomagnetic dipole moments can also be probed in \epem\ collisions
by studying gluon emission in top pair production.  In fact, the 
presence of such couplings significantly modify the gluon energy distribution
in $\epem\to t\bar tg$.  A two parameter fit to a Monte Carlo generated data 
sample for the gluon energy spectrum has been 
performed in Ref. \citenum{tgrnlc}, where a judiciously
chosen cut on the minimum gluon jet energy $E_g^{min}$ has been imposed.
This cut is necessary to identify the event as $t\bar tg$ and avoid the
infra-red singularities as well as avoiding contamination from the additional
gluon radiation off of the b-quarks in the final state.  The resulting $95\%$ 
C.L. search region in the chromomagnetic - chromoelectric ($f_2^g - f_4^g$) 
moment plane is
displayed in Fig. \ref{ttg} for the cuts and luminosities as indicated.  
We see that the obtainable bounds on $f_2^g$
are comparable to those of the LHC.

\vspace*{-0.5cm}
\nn
\vspace*{0.4mm}

\begin{figure}[t]
\centerline{
\psfig{figure=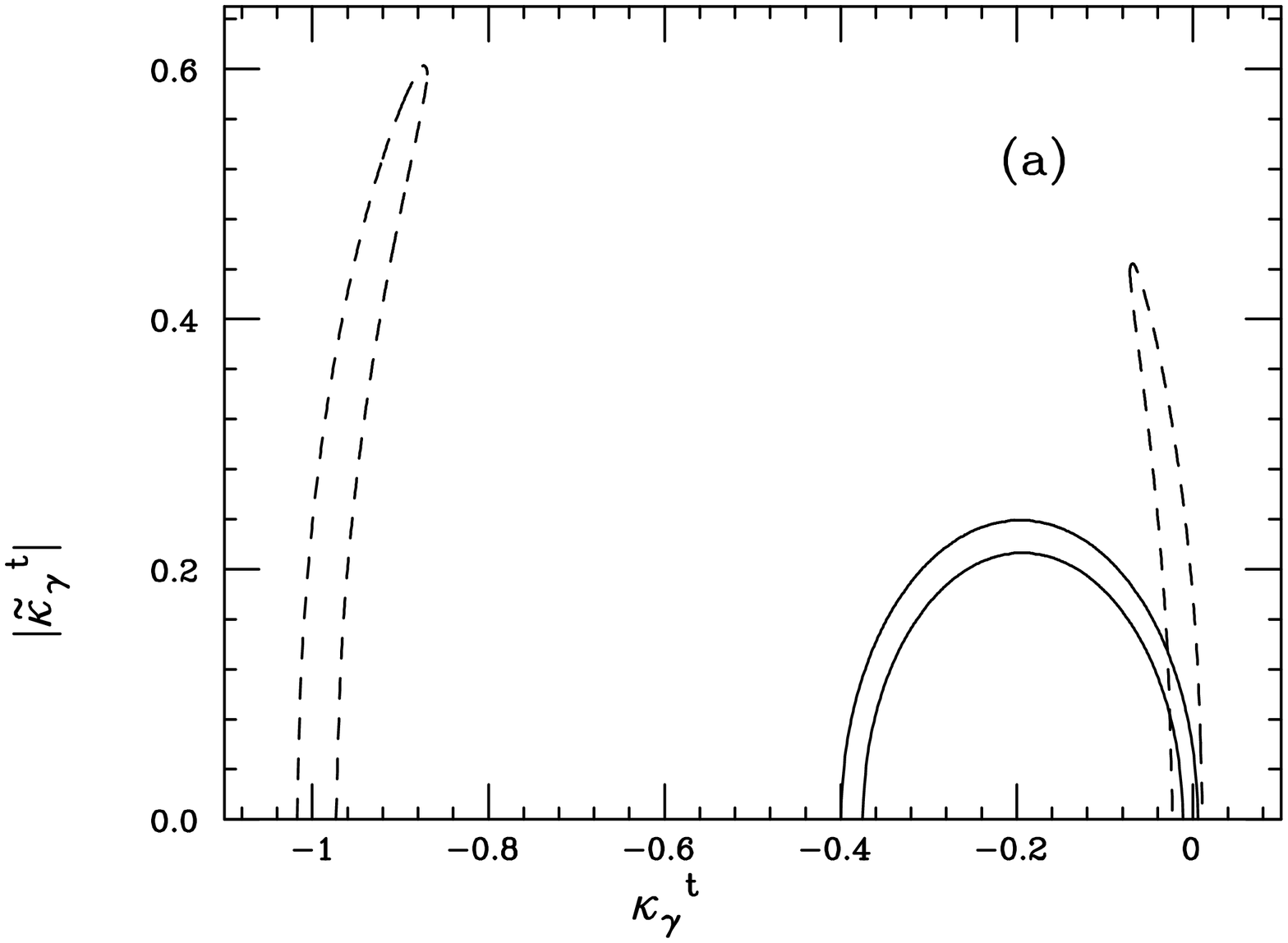,height=6.5cm,width=7cm,angle=-90}
\hspace*{-5mm}
\psfig{figure=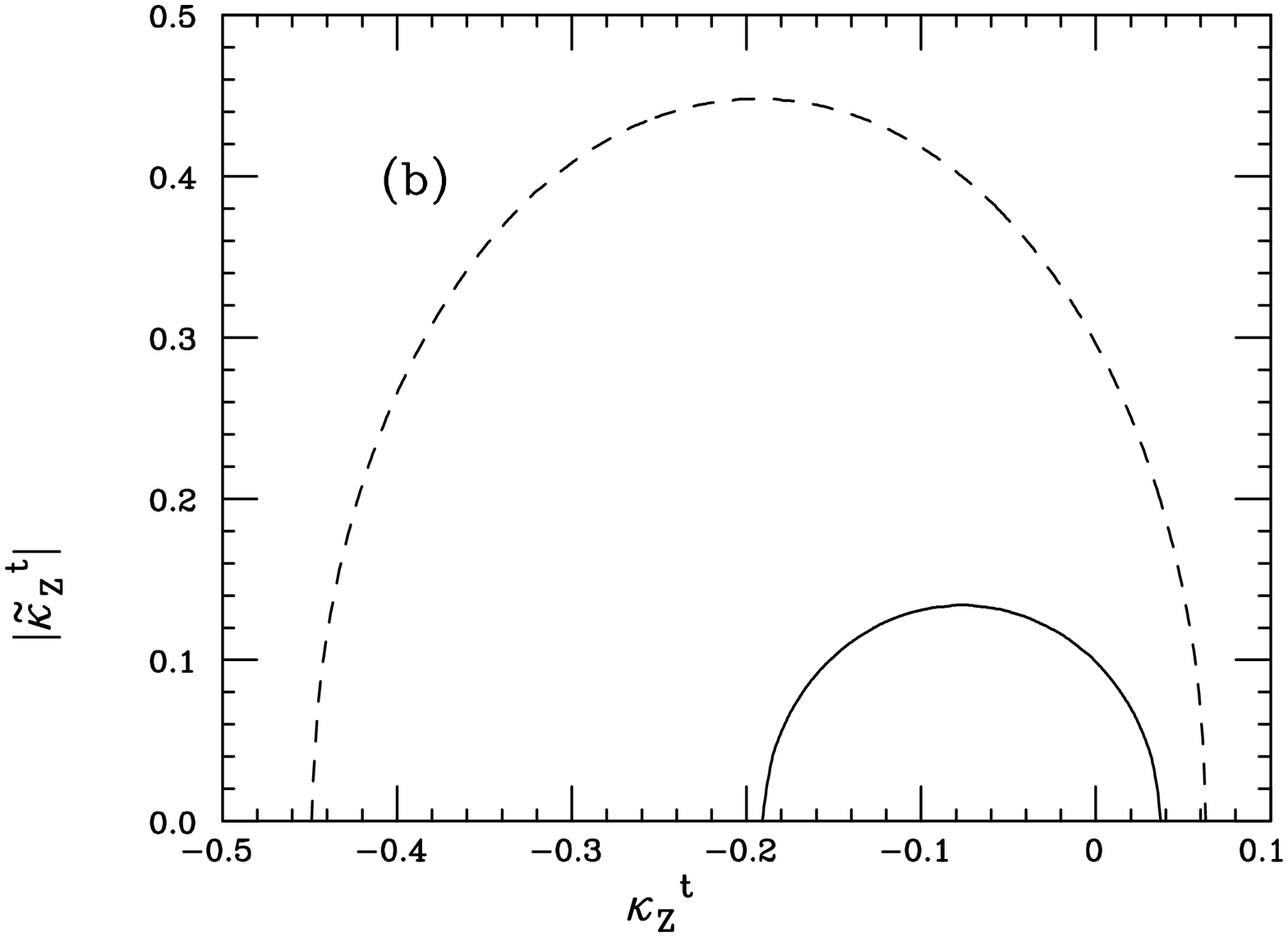,height=6.5cm,width=7cm,angle=-90}}
\vspace*{-0.5cm}
\fcaption{$95\%$ C.L. allowed regions in the $f_4^\gamma - f_2^\gamma$ 
and $f_4^Z - f_2^Z$ planes (denoted
here as $\tilde\kappa_{\gamma,Z} - \kappa_{\gamma,Z}$) from 
Ref. \citenum{tgrnlc}.
The allowed regions lie inside the dashed (solid) curves 
for a 500 (1000) GeV NLC assuming 50 (100) fb$^{-1}$.}
\label{ttgam}
\end{figure}

Of course, the presence of non-standard $t\bar t\gamma/Z$ couplings at the top
pair production vertex can also modify the shape of the gluon energy
spectrum in $\epem\to t\bar tg$.  Following the same Monte Carlo procedure as 
above\cite{tgrnlc} the $95\%$ C.L. regions are obtained, taking, for
simplicity, only two of the couplings to be simultaneously non-vanishing.
The results are presented in Fig. \ref{ttgam} in the $f_2^\gamma - f_4^\gamma$
and $f_2^Z - f_4^Z$ planes.  We see that the obtainable constraints on the
$t\bar t\gamma$ couplings are qualitatively stronger than those for
$t\bar tZ$.

\vspace*{-0.5cm}
\nn
\vspace*{0.4mm}

In a full analysis at the NLC one also needs to fold in the interplay between 
the possible form factors in both top-quark production and decay.  For the 
top-quark decay, we can write the $tbW$ three point function as
\begin{equation}
{\cal L}={g\over\sqrt 2}\bar t\left[
\gamma_\mu ( P_Lf^W_{1L}+P_Rf^W_{1R})
+{ig\sigma_{\mu\nu}q^\nu\over 2\sqrt 2m_t}( P_Lf^W_{2L}
+P_Rf^W_{2R})\right] b W^\mu + h.c. \,,
\end{equation}
where $P_{L,R}$ are the helicity projection operators.  In the SM, $f^W_{1L}=1$
and remaining form factors vanish.  In the analysis of Ref. \citenum{nlczdr},
all relevant observables for the process $\epem\to t\bar t$ 
are combined in a likelihood function, using a Monte Carlo generator which
includes the helicity information for $t\bar t(g)$ production calculated to 
${\cal O}(\alpha_s)$.  An essential ingredient for this analysis is the fact
that top-quark pairs are produced in an approximately unique spin configuration
in polarized \epem\ collisions.\cite{yael}
The resulting simultaneous bounds on non-standard top 
couplings are given in Table \ref{epemtotal} for 10 \infb\ at $\sqrt s=500$ GeV
with both unpolarized and $80\%$ left-polarized electron beams.  We see that
the overall sensitivity to possible deviations in these couplings is at the 
$10-15\%$ level.  This would correspond, for example, to a $t\bar tZ$ electric
dipole moment of $8\times 10^{-18}$ e-cm.

\begin{table}
\centering
\tcaption{Results from the global top-quark form factor analysis of 
Ref. \citenum{nlczdr} with a 10 ${\rm fb}^{-1}$ data sample at a 500 GeV NLC.}
\begin{tabular}{|c|c|c|c|} \hline\hline
Form Factor & SM Value & Limit & Limit \\ 
            & Lowest Order & $68\%$ C.L. & $68\%$ C.L.\\ \hline
$f^W_{1R}(P=0)$ & 0 & $\pm 0.13$ & $\pm 0.18$ \\
$f^W_{1R}(P=80\%)$ & 0 & $\pm 0.06$ & $\pm 0.10$ \\
$f_1^Z$ & $v_t$ & $v_t(1\pm 0.10)$ & $v_t(1\pm 0.16)$ \\
$f_3^Z$ & $a_t$ & $a_t(1\pm 0.08)$ & $v_t(1\pm 0.13)$ \\
$f_2^\gamma$ & 0 & $\pm 0.07$ & $~^{+0.13}_{-0.11}$ \\
$f_4^\gamma$ & 0 & $\pm 0.05$ & $\pm 0.08$ \\
$f_2^Z$ & 0 & $\pm 0.07$ & $\pm 0.10$ \\
$f_4^Z$ & 0 & $\pm 0.09$ & $\pm 0.15$\\
$\Im (f_4^Z)$ & 0 & $\pm 0.06$ & $\pm 0.09$ \\ \hline\hline
\end{tabular}
\label{epemtotal}
\end{table}

\begin{figure}[t]
\centerline{
\psfig{figure=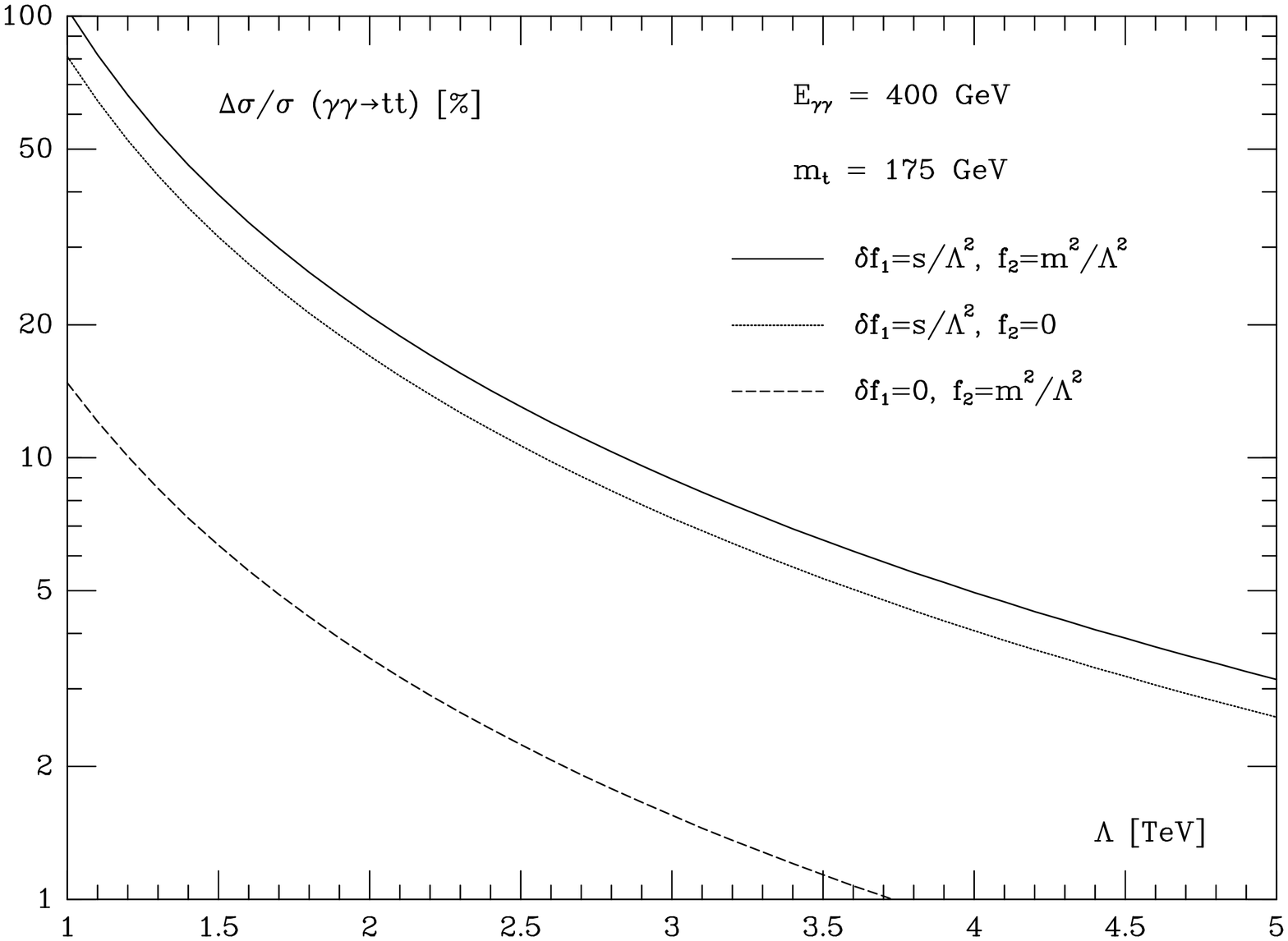,height=8cm,width=10cm,angle=-90}}
\vspace*{0.25cm}
\fcaption{Percentage change of the total cross section $\gg\to t\bar t$ from its
SM value for the various form factors as labeled, taking 
$\sqrt{s_{\gamma\gamma}}=400$ GeV.}
\label{gamgam}
\end{figure}

The possibility of turning the \epem\ collider into a \gg\ collider via
back-scattered laser beams can be exploited to measure the $t\bar t\gamma$
couplings independently of the $t\bar tZ$ couplings.  In this case,
the most general form of the differential cross section 
$d\sigma(\gamma\gamma\to t\bar t)/d\cos\theta$ for the $t\bar t\gamma$
couplings in Eq. (\ref{lagrange}) has been computed in Ref. \citenum{djouadi}.
The deviations in the total cross section as a function of the new physics
scale $\Lambda$ are presented in Fig. \ref{gamgam} for the cases (i)
$\delta f_1=s/\Lambda^2\,, f_2=m_t^2/\Lambda^2$, (ii) 
$\delta f_1=s/\Lambda^2\,, f_2=0$, (iii)
$\delta f_1=0\,, f_2=m_t^2/\Lambda^2$, where $\delta f_1$ represents the
shift from its SM value.  Here, a fixed $\gamma\gamma$ c.m. energy 
of $\sqrt{s_{\gamma\gamma}}=400$ GeV is assumed.  
Note that the effects of the non-standard couplings are found to be
slightly more pronounced than in the $\epem\to t\bar t$ 
production.\cite{djouadi}
Measurements of the total cross section alone with an
accuracy of $2\%$ could probe new interaction scales up to 10 TeV. In order
to compare the \gg\ reach with the results in Table \ref{epemtotal} we note
that with the above normalization taking $\Lambda=2.5 (5.0)$ TeV, corresponds to
$\delta f_1=\pm 0.025 (0.006)$ and $f_2=\pm 0.005 (0.001)$.

\vspace*{-0.5cm}
\nn
\vspace*{0.4mm}

\section{Exploring CP Violation with \boldmath$\gamma$ Beams}

The potential to polarize the Compton backscattered photon beam gives rise
to interesting ways of studying CP violating effects in $\gg\to t\bar t$.
The luminosity distributions of the backscattered photons for various
helicity combinations of the laser beam and initial electron beam display
the following properties:\cite{bauer}
in the case of $\lambda_e\lambda_\gamma=
-1$, the luminosity peaks at higher values of invariant mass with a maximum
at roughly $W_{\gg}/E_{ee}=80\%$, while for $\lambda_e\lambda_\gamma=+1$ the
spectrum is almost Gaussian like (for low values of the intrinsic photon
spread), peaking at lower energies.  These characteristics allow the
measurement of various CP-odd correlations.
For example, CP-odd asymmetries can be formed when the two $J=0$ amplitudes
\begin{eqnarray}
{\cal A}[\gg\to X(CP=+1)] & \sim & \vec\epsilon_1\cdot\vec\epsilon_2\,,\\
{\cal A}[\gg\to X(CP=-1)] & \sim & (\vec\epsilon_1\times\vec\epsilon_2)
\cdot\vec k \,,\nonumber
\end{eqnarray}
are non-vanishing, with $\vec\epsilon_{1,2}$ being the polarization of the
two photon beams and $\vec k$ the momentum carried by one of the photons
in the \gg\ c.m. frame.  Here, $X$ represents a non-CP eigenstate, such as,
$t\bar t$ in the case of non-vanishing $t\bar t\gamma$ electric dipole
moments, or $X$ can correspond to a Higgs boson with CP violating couplings
in multi-Higgs models.  The first example has been studied in Refs.
\citenum{choi,baek,poulose}.

Choi and Hagiwara\cite{choi} have investigated a polarization asymmetry
which isolates the electric dipole moment contribution by taking the
difference of $t\bar t$ event rates at $\chi=+\pi/4$ and $\chi=-\pi/4$,
where $\chi$ is the angle between the directions of maximum linear
polarization of the two laser beams in the \epem\ (or $e^-e^-$) c.m. frame.
For $\sqrt s=500$ GeV with 20 \infb, they find that this asymmetry can
probe top electric dipole moments down to $e|\Re f_4^\gamma|/2m_t<
1.16\times 10^{-17}$ e-cm for $m_t=170$ GeV. Baek \etal,\cite{baek} perform
a similar analysis and find $|\Re f_4^\gamma|< 0.16$, for the same machine
parameters.  Note that these results are comparable to the potential
constraints listed in Table \ref{epemtotal} obtained from top decay spin
correlation analyses in $\epem\to t\bar t$.  
Poulose and Rindani\cite{poulose}
have obtained similar results by examining the charge and combined charge
and forward-backward asymmetries.  Folding in the full photon energy and
luminosity distributions they obtain the $90\%$ C.L. bounds on $\Im f_4^\gamma$
shown in Table \ref{imf4} for various helicity combinations as listed and
for electron beam energies of 250 GeV.  These limits are shown to improve by
an order of magnitude for 500 GeV electron beams.

\begin{table}
\centering
\tcaption{$90\%$ C.L. bounds on $\Im f_4^\gamma$ in units of $10^{-16}$ e-cm,
for $m_t=174$ GeV, electron beam energy of 250 GeV, and 20 \infb.  From Ref.
\protect{\citenum{poulose}}.}
\begin{tabular}{|c|c|c|c|c|} \hline\hline
$\lambda^1_e$ & $\lambda^2_e$ & $\lambda^1_\gamma$ & $\lambda^2_\gamma$ &
$\Im f_4^\gamma$ \\
 & & & & ($10^{-16}$ e-cm) \\ \hline
$-0.5$ & $-0.5$ & $-1$ & $-1$ & 2.76 \\
$-0.5$ & $-0.5$ & $1$ & $-1$ & 0.23 \\
$-0.5$ & $-0.5$ & $1$ & $1$ & 0.54 \\
$0.5$ & $-0.5$ & $-1$ & $-1$ & 2.31 \\
$0.5$ & $-0.5$ & $1$ & $-1$ & 1.03 \\
$0.5$ & $-0.5$ & $1$ & $1$ & 1.12 \\
\multicolumn{4}{|c|} {Unpolarized} & 1.19\\ \hline\hline
\end{tabular}
\label{imf4}
\end{table}

\begin{figure}[t]
\centerline{
\psfig{figure=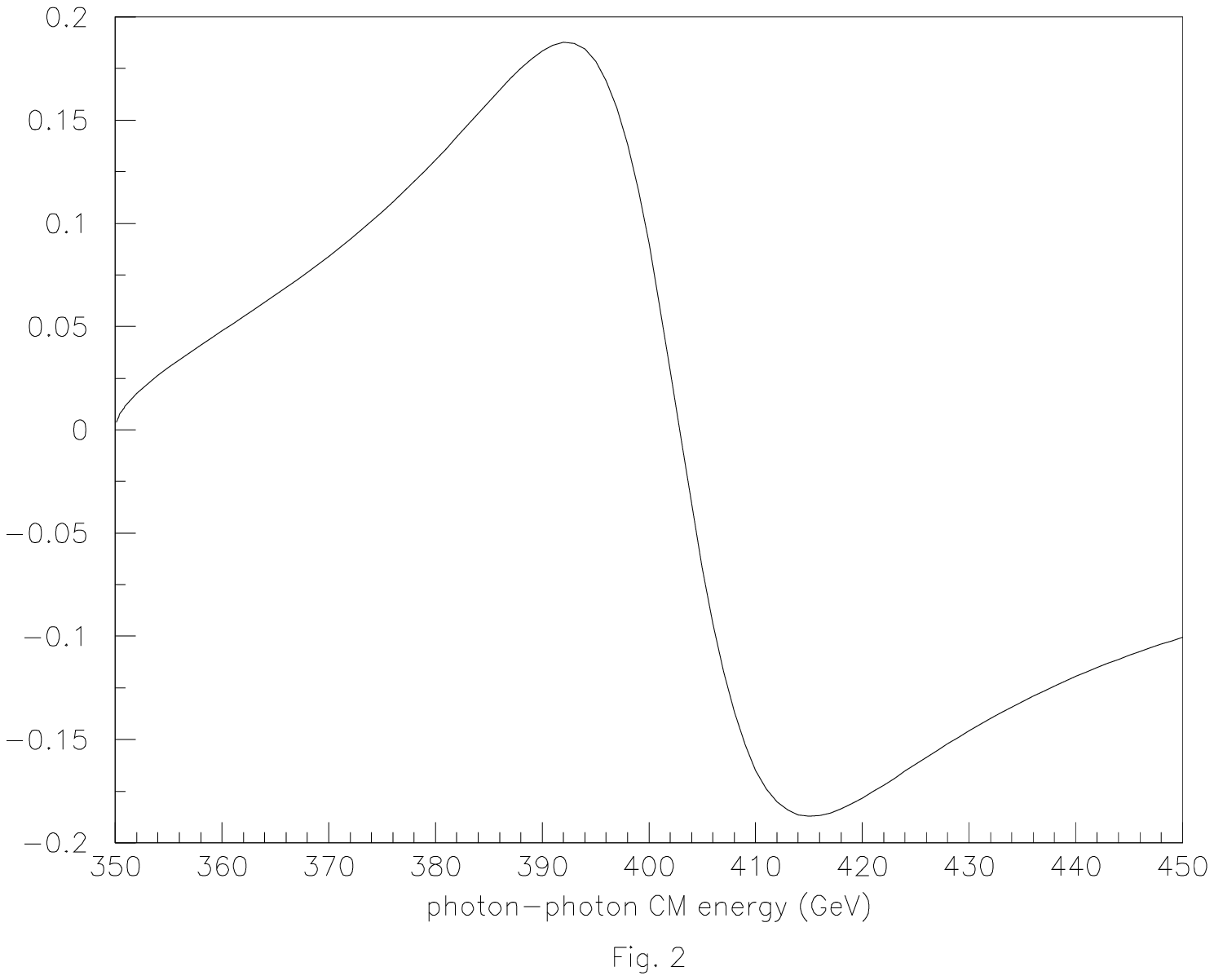,height=6cm,width=10cm,angle=0}}
\vspace*{1.25cm}
\fcaption{Longitudinal polarization asymmetry described in the text as a
function of the \gg\ c.m. energy.}
\label{anl2}
\end{figure}

CP violation studies in $\gg\to H\to t\bar t$ with unpolarized beams have
been performed in Ref. \citenum{abb}, using triple-product correlations and
a variety of polarization asymmetries in top semi-leptonic decays.  These
authors examine the case of two-Higgs-doublet models which contain explicit
CP violation in both the Yukawa couplings and in the Higgs potential.  A
typical result is given by the longitudinal polarization asymmetry defined
by $\hat{\bf k}\cdot({\bf s}_+ - {\bf s}_-)$, where ${\bf k}$ is the 
momentum of the $t$-quark and ${\bf s}_+\,, {\bf s}_-$ are the spin operators
of $t$ and $\bar t$, respectively.  This observable is displayed in Fig.
\ref{anl2} as a function of the \gg\ c.m. energy, where the mass of the
lightest neutral Higgs boson is taken to be 400 GeV, the ratio of
vacuum expectation values of the two doublets is $\tan\beta\equiv v_2/v_1=1$,
and $m_t=175$ GeV.
We see that the asymmetry grows large near the resonance region.  The
other observables discussed in this reference display similar characteristics
and could provide a window into the CP properties of extended Higgs sectors.

\vspace*{-0.5cm}
\nn
\vspace*{0.4mm}

\section{Flavor Changing Top Couplings}

Another interesting possibility is to study the flavor changing top-quark
vertices $tq\gamma/Z$ where $q=c$ or $u$ at lepton and photon colliders.
The expectations in the SM (and simple extensions thereof, such as 
supersymmetry)\cite{ehs} for these 
couplings are too small to be observed and thus any experimental evidence for
their existence implies new interactions.  Here, we consider
only the lowest dimension CP-conserving operators in the effective Lagrangian
involving the anomalous top-quark couplings,
\begin{equation}
{\cal L}^{eff} = {e\over\Lambda}\kappa_\gamma\bar t\sigma_{\mu\nu}cF^{\mu\nu}
+{g\over 2c_w}\bar t\left[g_{ZL}\gamma_\mu P_L+g_{ZR}\gamma_\mu P_R + 
{i\sigma_{\mu\nu}q^\nu\over m_t}\kappa_Z\right]qZ^\mu +h.c. \,,
\end{equation}
where $\Lambda$ is the cutoff of the effective theory, and $P_{L,R}$ are
the usual helicity projection operators.  CDF has placed the $95\%$ C.L. 
bounds on the flavor
changing neutral current (FCNC) decays of the top-quark of $B(t\to cZ+uZ)<33\%$
and $B(t\to c\gamma+u\gamma)<2.9\%$.  The latter limit implies
$\kappa_\gamma/\Lambda<0.73/{\mbox {TeV}}$.  An extensive 
analysis\cite{hanpp} 
has shown that the LHC can improve on these constraints by directly
searching for the FCNC decays and probe the pertinent anomalous couplings
to the level of $\kappa_\gamma/\Lambda\simeq 0.01/{\mbox {TeV}}$ and
$g_{tc}=\sqrt{g^2_{ZL}+g^2_{ZR}}\simeq 0.02$ which corresponds
to a branching fraction of $B(t\to qZ)<2\times 10^{-4}$.

\begin{figure}
\vspace*{13pt}
\vspace*{2.8truein}      
\includegraphics{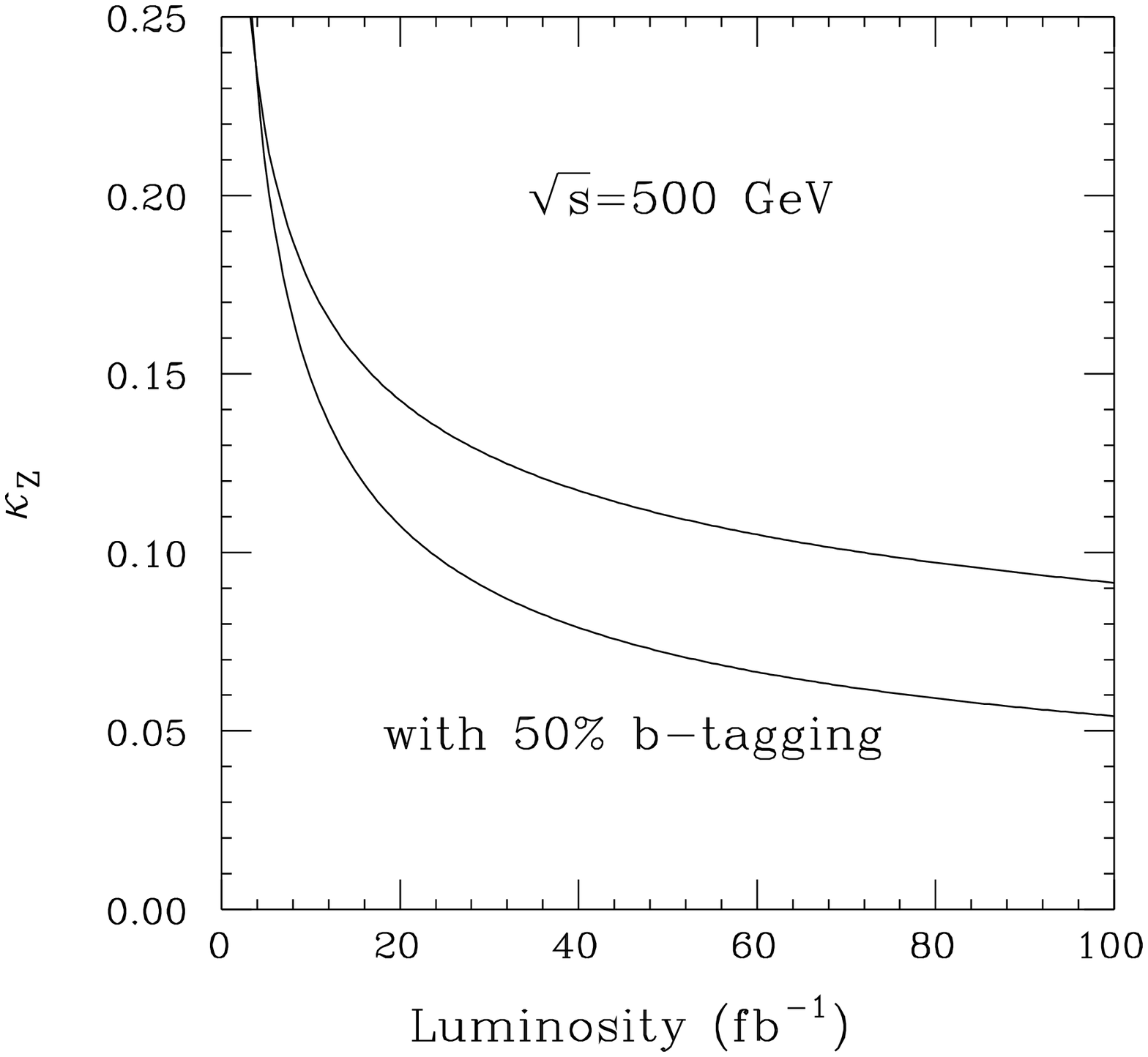}
\vspace{1.25cm}
\fcaption{Expected $95\%$ C.L. bounds on $\kappa_Z$ as a function of luminosity
at a 500 GeV NLC, with and without a $50\%$ b-tagging identification
efficiency as indicated.}
\label{halftev}
\end{figure}

The NLC can probe these  couplings in the direct production vertex via
the reaction $\epem\to t\bar c+\bar tc$.  In this case, the cross sections
are found\cite{hannlc} to be large and the signature clean with a single
hard jet in one hemisphere and a top-quark decaying to $bW\to b\ell\nu_\ell$
in the other.  These kinematic characteristics allow for a clear separation
of the signal from the $WW$ pair background by imposing a set of idealized cuts.
Figure \ref{halftev} displays the resulting $95\%$ C.L. bounds obtainable on 
$\kappa_Z$ from this analysis,
with and without imposing a $50\%$ b-tagging identification efficiency,
as a function of luminosity for $\sqrt s= 500$ GeV.  The expectations for
constraints on $\kappa_\gamma$ are comparable to those above for the LHC,
\ie, $\kappa_\gamma/\Lambda\simeq 0.01/{\mbox {TeV}}$.

The radiative flavor changing couplings can also be tested at a \gg\ collider
via $t\bar q+\bar tq$.  This has been investigated in Ref. \citenum{abraham},
with the result
\begin{equation}
\sigma(\gg\to t\bar c+\bar tc\to b\ell^+\nu\bar c+\bar b\ell^-\bar\nu c)
= \left[ {\kappa_\gamma/\Lambda\over 0.16/{\mbox {TeV}}}\right]^2
76.4\, {\mbox {fb}} \,.
\end{equation}
This corresponds to an expected search reach of $\kappa_\gamma/\Lambda
< (0.05/{\mbox TeV})/\sqrt{{\cal L}/10\,\infb}$ for a fixed \gg\ c.m. energy
of 400 GeV.

\section{Summary}

We have presented an overview of the ability of lepton and photon colliders
to probe CP conserving and violating anomalous couplings of the top-quark.  
In particular, some advantages of a \gg\ collider
over the \epem case
is that the $t\bar t\gamma$ three-point function can be isolated from that for
$t\bar tZ$, the $t\bar t$ production cross section is larger, and the
potential for linear photon polarization provides some simple probes of
CP violation.  When comparisons can be made, we see that both types of
machines can yield comparable, if not better, constraints than those obtainable
in hadronic collisions.

\def\MPL #1 #2 #3 {Mod. Phys. Lett. {\bf#1},\ #2 (#3)}
\def\NPB #1 #2 #3 {Nucl. Phys. {\bf#1},\ #2 (#3)}
\def\PLB #1 #2 #3 {Phys. Lett. {\bf#1},\ #2 (#3)}
\def\PR #1 #2 #3 {Phys. Rep. {\bf#1},\ #2 (#3)}
\def\PRD #1 #2 #3 {Phys. Rev. {\bf#1},\ #2 (#3)}
\def\PRL #1 #2 #3 {Phys. Rev. Lett. {\bf#1},\ #2 (#3)}
\def\RMP #1 #2 #3 {Rev. Mod. Phys. {\bf#1},\ #2 (#3)}
\def\ZPC #1 #2 #3 {Z. Phys. {\bf#1},\ #2 (#3)}
\def\IJMP #1 #2 #3 {Int. J. Mod. Phys. {\bf#1},\ #2 (#3)}
\nonumsection{References}


\begin{thebibliography}{99}

\bibitem{topmass}
B. Klima, talk given at the 
{\it 12th Les Recontre de Physique de la Vallee d'Aoste:
Results and Perspectives in Particle Physics}, La Thuile, Italy, March 1998.
%
\bibitem{ewwg}
The LEP Collaborations, the LEP Electroweak Working Group, and the SLD
Heavy Flavor Group, CERN-PPE/97-154 (1997).
%
\bibitem{htt}
For a list of references, see, J.L. Hewett, T. Takeuchi, and S. Thomas in
{\it Electroweak Symmetry Breaking and New Physics at the TeV Scale},
ed. T. Barklow \etal, (World Scientific, Singapore 1996), hep-ph/9603391.
%
\bibitem{escrib}
R. Escribano and E. Masso, \PLB B301 419 1993 , and \NPB B429 19 1994 .
%
\bibitem{bartl}
A. Bartl, \etal, \NPB B507 35 1997 .
%
\bibitem{columbia}
R. Martinez \etal, hep-ph/9709478.
%
\bibitem{jlhtgr}
J.L. Hewett and T.G. Rizzo, \PRD D49 319 1994 .
%
\bibitem{cleo}
M.S. Alam \etal, CLEO Collaboration, \PRL 74 2885 1995 .
%
\bibitem{sally}
S. Dawson and G. Valencia, \PRD D53 1721 1996 .
%
\bibitem{tgr}
D. Atwood, A. Kagan, and T.G. Rizzo, \PRD D52 6264 1995 ; T.G. Rizzo,
\PRD D53 6218 1996 ; K. Cheung, \PRD D53 3604 1996 ; P. Haberl, O.
Nachtmann, and A. Wilch, \PRD D53 4875 1996 . 
%
\bibitem{tgrlhc}
T.G. Rizzo in {\it New Directions
for High-Energy Physics}, Snowmass, CO, 1996, ed. D.G. Cassel, hep-ph/9609311.
\bibitem{tgrnlc}
T.G. Rizzo, \PRD D50 4478 1994 , and in {\it New Directions
for High-Energy Physics}, Snowmass, CO, 1996, ed. D.G. Cassel, hep-ph/9610373.
%
\bibitem{nlczdr} 
{\it Physics and Technology of the Next Linear Collider}, BNL Report BNL
52-502, 1996; R. Frey in {\it Proceedings of the Workshop on Physics and
Experiments with Linear Colliders}, Iwate, Japan, 1995, hep-ph/9606201;
C. Schmidt, hep-ph/9504434; C.R. Schmidt, \PRD D54 3250 1996.
%
\bibitem{yael}
S. Parke and Y. Shadmi, \PLB B387 199 1996 .
%
\bibitem{djouadi}
A. Djouadi, in {\it Proceedings of the Workshop on $e^+e^-$ Collisions at 500
GeV: the Physics Potential}, ed. P. Zerwas, Report DESY-92-123B; A. Djouadi,
J. Ng, and T.G. Rizzo, in 
{\it Electroweak Symmetry Breaking and New Physics at the TeV Scale},
ed. T. Barklow \etal, (World Scientific, Singapore 1996), hep-ph/9504210.
%
\bibitem{bauer}
See, for example, D. Bauer, \IJMP A11 1637 1996 ; D.L. Borden in
{\it Proceedings of the Workshop on Physics and Experiments with Linear
\epem\ Colliders}, ed. F.A. Harris \etal, (World Scientific, Singapore 1993).
%
\bibitem{choi}
S.Y. Choi and K. Hagiwara, \PLB B359 369 1995 .
%
\bibitem{baek}
M.S. Baek \etal, \PRD D56 6835 1997 .
%
\bibitem{poulose}
P. Poulose and S.D. Rindani, hep-ph/9709225.
%
\bibitem{abb}
H. Anlauf, W. Bernreuther, and A. Brandenburg, \PRD D52 3803 1995 , Erratum
{\bf D53}, 1725 (1996).
%
\bibitem{ehs} See, for example, G. Eilam, J.L. Hewett, and A. Soni,
\PRD D44 1473 1991 .
%
\bibitem{cdftop}
G. Chiarelli (CDF Collaboration), in {\it Proceedings of the 32nd Rencontres de
Moriond: QCD and High Energy Hadronic Interactions}, Les Arcs, France, March
1997, Fermilab-CONF-97/143-E.
%
\bibitem{hanpp}
T. Han \etal, \PRD D55 7241 1997 ; T. Han, R. Peccei, and Z. Zhang, 
\NPB B454 527 1995 .
%
\bibitem{hannlc}
T. Han and J.L. Hewett, in preparation.
%
\bibitem{abraham}
K.J. Abraham \etal, hep-ph/9707476.

\end{thebibliography}
\end{document}